\documentclass[a4paper,12pt]{article}

\usepackage{authblk}
\usepackage[margin=1in]{geometry}
\usepackage{bm}
\usepackage{amssymb}
\usepackage{amsfonts}
\usepackage{mathrsfs}
\usepackage{amsmath}
\usepackage{amsthm}
\usepackage{xcolor}
\usepackage{bbold}
\usepackage{braket}
\usepackage{physics}
\usepackage{cases}
\usepackage{float}
\usepackage{caption}
\usepackage{subcaption}
\usepackage{graphicx}
\usepackage{wrapfig}

\newcommand{\be}{\begin{equation}}
\newcommand{\ee}{\end{equation}}
\newcommand{\bea}{\begin{eqnarray}}
\newcommand{\eea}{\end{eqnarray}}
\def\bml{\begin{subequations}}
\def\blea{\bml\begin{eqnarray}}
\def\eml{\end{subequations}}
\def\elea{\end{eqnarray}\eml}

\newtheorem{theorem}{Theorem}[section]

\newtheorem{prop}[theorem]{Proposition}
\theoremstyle{definition}

\theoremstyle{remark}

\usepackage{titlesec}

\titleformat*{\section}{\small\bfseries}

\usepackage{setspace}
\title{The sufficiently trapped surface}
\author{Eleni-Alexandra Kontou \thanks{eleni.kontou@kcl.ac.uk}}
\affil{Department of Mathematics, King’s College London,\\Strand, London, WC2R 2LS, United Kingdom}
\date{}

\begin{document}

\maketitle

\begin{abstract}
Roger Penrose introduced the concept of the trapped surface: a spacelike hypersurface where the two null normals have negative expansion. The trapped surface along with the null convergence condition leads to null geodesic incompleteness. If an event horizon forms, the trapped surface is also always behind it, providing evidence for the weak cosmic censorship conjecture. When the null convergence condition is violated, as in the case of semiclassical gravity, trapped surfaces lose these guarantees. A generalized notion, the sufficiently trapped surface, accommodates weaker energy conditions consistent with quantum fields. This concept restores key roles in singularity and area theorems and continues to support the weak cosmic censorship conjecture.\\
\textit{Essay written for the Gravity Research Foundation 2026 Awards for Essays on Gravitation and awarded ``honorable mention''.}
\end{abstract}

\newpage

\section{Introduction}

Often, the defining feature of a black hole is the presence of an event horizon, the boundary of the region from which no signals can reach future infinity. By its very nature, the event horizon is a global and inherently \textit{teleological} object, as its location depends on the entire future evolution of spacetime.

For many purposes, it is more practical to work with quasi-local notions that capture the structure of black holes without reference to the distant future. In 1965, Penrose introduced one of the most influential concepts in gravitational physics: the trapped surface \cite{penrose1965gravitational}. A trapped surface is a spacelike hypersurface whose two null normals both have negative expansion\footnote{Throughout this essay, trapped surfaces are assumed to be closed, i.e., compact and without boundary.}. Equivalently, the cross-sectional area of any congruence of causal geodesics orthogonal to the surface decreases in any direction.

Trapped surfaces play a central role in the singularity theorem of Penrose \cite{penrose1965gravitational}. This theorem can be understood as relying on three key ingredients: (i) an initial condition, (ii) an energy condition, and (iii) a causality condition. The initial condition is precisely the existence of a trapped surface. The relevant energy condition is the null convergence condition (NCC),
\begin{equation}
\label{eqn:NCC}
R_{\mu \nu} U^\mu U^\nu \geq 0 \,,
\end{equation}
for all null vector fields $U^\mu$. Via the Einstein equations, this can also be interpreted as the null energy condition.

Together, the initial and energy conditions ensure the formation of focal points along null geodesics orthogonal to the surface. More precisely, these are points where Jacobi fields vanish, indicating the convergence of geodesic congruences. The causality condition — typically global hyperbolicity — prevents such geodesics from extending beyond these focal points, implying that they must be incomplete. In this context, geodesic incompleteness is taken as the definition of a spacetime singularity.

When an event horizon is present, an additional structure emerges. Under the assumption of strong asymptotic predictability, Hawking showed \cite{Hawking:1971vc} that trapped surfaces must lie entirely within the event horizon, a result that relies on the NCC. This has important implications for the weak cosmic censorship conjecture, originally proposed by Penrose \cite{Penrose:1969pc} which asserts that singularities are always hidden behind event horizons and are therefore not visible to distant observers. If the NCC holds and an event horizon forms, the singularity is necessarily concealed, lending support to the conjecture.

Penrose further developed these ideas through the Penrose inequality \cite{PenroseNakedSing1973}. Consider a spacetime containing a trapped surface: Assuming weak cosmic censorship, an event horizon forms outside it. The NCC, together with the existence of an event horizon, are the conditions for Hawking’s area theorem \cite{Hawking:1971vc}, which states that the area of the event horizon is non-decreasing. If, in addition, the spacetime settles down to a Kerr solution, one can relate the horizon area to the Arnowitt–Deser–Misner (ADM) mass \cite{Arnowitt:1960es}. This leads to the inequality
\begin{equation}
\label{eqn:PI}
m_{\mathrm{ADM}} \geq \sqrt{\frac{A_{\min}(\Sigma)}{16\pi}} \,,
\end{equation}
where $A_{\min}(\Sigma)$ is the infimum of the areas of all spacelike surfaces enclosing the trapped surface $\Sigma$. A violation of this inequality would indicate a breakdown of weak cosmic censorship and the formation of a naked singularity.

The situation changes dramatically in the presence of quantum effects. Hawking’s discovery \cite{Hawking:1975vcx} that black holes emit thermal radiation implies that they can lose mass and evaporate when backreaction is taken into account. A key feature of this process is the violation of the null energy condition—and hence the NCC—by quantum fields (e.g. \cite{PhysRevD.56.936, levi2016versatile}). Such violations are inherently quantum, as quantum fields are known to evade all classical pointwise energy conditions \cite{Epstein1965}. This has far-reaching consequences. The classical singularity and area theorems no longer directly apply, and trapped surfaces need not lie within the event horizon. In some cases, they may even lie outside it; for example, in evaporating Vaidya spacetimes, the apparent horizon eventually lies outside the event horizon \cite{Hiscock:1980ze}.

Thus, in semiclassical gravity, the notion of a trapped surface is no longer sufficient to establish the existence of singularities, determine the behavior of event horizons, or support the weak cosmic censorship conjecture. This motivates the introduction of the concept of a sufficiently trapped surface, a refinement of Penrose’s original definition. The precise strengthening required depends on the extent to which negative energy is permitted by the matter model. Exploring this generalized notion and its implications is the central aim of this essay.

\section{The formation of focal points}

The formation of a focal point can be examined either using the Raychaudhuri equation or directly, using the index form method \cite{o1983semi}. Here we will consider the latter. The test to determine the formation or not of a focal point for causal geodesics is the sign of the Hessian of the action integral $E$ of the geodesic. For timelike geodesics the relevant integral is the second variation of the length of the geodesic, also called the index form.  

Let us consider a causal curve \(\gamma : [0, \ell] \rightarrow M\) affinely parameterized by \(\lambda\) emanating normally from a spacelike hypersurface $P$. Then the action integral \(E\) is defined as
\be
E[\gamma] := \frac{1}{2}\int_{0}^{\ell} g(\gamma'(\lambda), \gamma'(\lambda))d\lambda \,.
\ee
Now we vary the curve in a transverse direction, parametrized by $s$. That leads to a family of curves $\gamma_s(\lambda):=\zeta(\lambda, s)$. The tangent and the transverse vector fields are defined as $U_\mu=\gamma'(\lambda)$ and $V_\mu=\partial \gamma_s/\partial s |_{s=0}$. The first variation of  $E[\gamma_s]$ is zero for geodesics, while its second variation is
\be
	\label{eq:hessian}
	\mathcal{H}[V]\equiv \frac{\partial^2E[\gamma_s]}{\partial s^2}\Big\vert_{s = 0} = 
	\int_{0}^{\ell} \left[(\nabla_UV_{\mu})(\nabla_UV^{\mu}) + R_{\mu\nu\alpha\beta}U^{\mu}V^{\nu}V^{\alpha}U^{\beta}\right] d\lambda -U^{\mu} \nabla_V V_{\mu} \Big\vert_0^{\ell}\,.
\ee
What is left is to choose a convenient $V$. Let \(e_i\) with \(i = 1, \ldots, n - 2\) be an orthonormal basis of \(T_{\gamma(0)}P\), and parallel transport it along \(\gamma\) to generate \(\{E_i\}_{i = 1, \ldots, n-2}\). Then, take \(f\) a smooth function with \(f(0) = 1\) and \(f(\ell) = 0\). Calculating the sum of Hessians for all $fE_i$ gives
\begin{equation}
	\label{eq:hessian-averagded}
	\sum_{i=1}^{n - 2}\mathcal{H}\left[fE_i\right] = - \int_{0}^{\ell} \left((n - 2)f'(\lambda)^2 - f(\lambda)^2R_{\mu\nu}U^{\mu}U^{\nu}\right) d\lambda - (n - 2)U_{\mu}H^{\mu}\Big\vert_{\gamma(0)},
\end{equation}
where $H^{\mu}$ is the mean curvature vector field of $P$. 

Now we can apply the condition for the formation of a focal point. If
\be
\mathcal{H}[V] \geq 0 \,,
\ee
for any $V$ then there is a focal point to $P$ along $\gamma$ \cite{o1983semi}. Using Eq.~\eqref{eq:hessian-averagded} this condition becomes
\begin{equation}
	\label{eq:fp-criteria}
		\int_0^\ell \big((n -2)f'(\lambda)^2 - f(\lambda)^2 R_{\mu \nu} U^\mu U^\nu \big)d\lambda \le -(n -2) U_\mu H^\mu \big|_{\gamma(0)} \,.
		\end{equation} 

We note that the term on the right hand side is a boundary term that characterizes the surface $P$. An important quantity associated with $P$ are the {\textit outward/inward null expansions} $\theta_\pm$. In a spacetime, the sign of the null expansions indicates the convergence and divergence of past and future directed null geodesics. Let $U_\pm$ be the future directed null normal vector fields along $P$. 

To connect this with Eq.~\eqref{eq:fp-criteria}, let the outward null direction be $U^\mu$. Then $U_\mu H^\mu \big|_{P} =  \theta_+$ on $P$. Outer and inner trapped surfaces are characterized by the inequalities $\theta_+ <0$ and $\theta_- < 0$ respectively. Thus $P$ is an outer trapped surface if $U_\mu H^\mu \big|_{P} <0$. 

A classical focusing theorem is then straightforward to prove using the condition \eqref{eq:fp-criteria}. Assuming (i) the NCC \eqref{eqn:NCC} and (ii) that $P$ is trapped surface, the only thing that remains is to pick a convenient function $f$. Picking a linear function $f(\lambda)=1-\lambda/\ell$ we have a focal point formed for 
\be
\ell\leq \frac{1}{\big|U_\mu H^\mu \big|_{P}} \,.
\ee
This result shows that with the NCC we have the formation of a focal point for any negative value of $U_\mu H^\mu \big|_{P}$, thus the requirement of an (outer) trapped surface is sufficient. 

\section{Violation of the null convergence condition}

In spacetimes where the NCC is violated the previous argument does not hold. We can deduce a weaker condition for the argument to proceed. However the argument only holds for a condition of the form
\be
\int_0^\ell  f(\lambda)^2 R_{\mu \nu} U^\mu U^\nu d\lambda \geq (n-2) \|f'(\lambda)\|^2 \,.
\ee
This condition reduces to the geometric form of the half-averaged null energy condition for a particular choice of the function $f$ \cite{Kontou:2023ntd}
\be
\int_0^\ell  f(\lambda)^2 R_{\mu \nu} U^\mu U^\nu d\lambda \geq 0 \,.
\ee

This condition is weaker than the NCC, but its physical form (the one with the stress energy tensor) is easily violated by quantum fields \cite{Ford:1995gb}, and thus it does not hold during black hole evaporation. 

We can then consider a class of more general conditions, inspired by quantum energy inequalities. Quantum energy inequalities are restrictions on the duration and magnitude of negative energy densities in the context of quantum field theory. Unlike the original energy conditions that are often imposed as assumptions to theorems, quantum energy inequalities are derived rigorously from the renormalized stress energy tensor for different quantum fields. They were introduced by Ford \cite{Ford:1978qya} and have been derived for multiple (mainly free) fields in flat and curved spacetimes (see \cite{Kontou:2020bta} for references). Their general form for a causal vector field $U^\mu$ is 
\be
 \int_{-\infty}^\infty dt \,\langle T_{\mu \nu} U^\mu U^\nu (\gamma(t)) \rangle_\omega g(t)^2 \geq  -\mathcal{Q}[g] \,,
\ee
where $g$ is a real test function of compact support, $\omega$ is the state of interest and the bound $\mathcal{Q}[g]$ is (sometimes under some further assumptions) state independent. The integral is usually over a timelike curve.

The bound $\mathcal{Q}[g]$ for scalar fields \cite{Eveson:2007ns} but also the electromagnetic field \cite{Fewster:2003ey} and the Dirac field \cite{Dawson:2006py} becomes a sum over constants and the $L^2$ norms of $g$ and its derivatives $Q_0\|g\|^2+Q_1\|g'\|^2+ ...$. This can always be written as the sum of the zeroth and the highest derivative terms by modifying the constants.

To translate the quantum energy inequalities into statements about spacetime curvature, we use the semiclassical Einstein equation as for classical energy conditions we use the Einstein equation
\be
G_{\mu \nu}=8\pi \langle T_{\mu \nu}^\text{ren} \rangle_\omega \,,
\ee
where $\langle T_{\mu \nu}^\text{ren} \rangle_\omega$ is the renormalized stress-energy tensor over a state $\omega$. A solution of this equation consists of a metric and a state. The semiclassical Einstein equation assumes that quantum fields self-consistently generate classical curvature, an assumption that can only have a limited regime of validity. That is the regime of low (classical) curvature but where quantum effects become important, for example the black hole horizon. 

Considering these two assumptions, the form of quantum energy inequalities and the semiclassical Einstein equation, we can suggest a curvature condition that replaces the NCC and could be obeyed by quantum fields
\be
\label{eqn:qeiinsp}
\int_{-\infty}^\infty d\lambda \, R_{\mu \nu}U^\mu U^\nu g^2(\lambda) \geq -Q_0 \| g\|^2-Q_m \|g^{(m)}\|^2 \,,
\ee
where the integral is over an affinely parametrized null geodesic \footnote{A complication that we do not discuss in detail here, is that quantum energy inequalities over finite null segments do not generally have finite lower bounds \cite{Fewster:2002ne}. Thus further assumptions are needed \cite{Freivogel:2018gxj,Fliss:2021phs}.}. The constant $m$ is a positive integer and $Q_0$ and $Q_m$ are non-negative constants. 

To use such a condition for a focusing theorem we can attempt to apply it directly to the condition for the formation of a focal point \eqref{eq:fp-criteria}. However, there is an important obstacle: the function $f$ in that condition has boundary conditions $f(0)=1$ and $f(\ell)=0$, while the function $g$ is a function of compact support. Thus, the condition cannot be applied immediately. Instead we can write $g=\phi f$ where $\phi$ is a function with $\phi(0)=0$ and $\phi(\ell)=1$. That leaves a gap in the estimate, for the segment of the geodesic where $f=1$. Assuming this segment is $[0,\ell_0]$ we can estimate that with a pointwise condition $R_{\mu \nu}U^\mu U^\nu \geq \rho_0$, where $\rho_0$ could be positive or negative. This still allows for violation of the NCC, and assuming $\ell \gg \ell_0$ the estimation of most of the negative energy is from the condition \eqref{eqn:qeiinsp}.

Applying these two conditions allows us to find when a focal point is formed. Specifically, this occurs when
\be
\label{eqn:trapqei}
U_\mu H^\mu \big|_{P} < -\nu(Q_0,Q_m,\rho_0, \ell_0, \ell) \,.
\ee
The positive function $\nu$ gives the required mean normal curvature to ensure the formation of a focal point. This depends primarily on how much negative energy is allowed. The exact expressions for $\nu$ for some specific functions are given in \cite{Fewster:2019bjg}. This result shows that a focusing theorem can hold in semiclassical gravity. However, a trapped surface alone is no longer sufficient to ensure geodesic focusing. Rather we need what we now call a sufficiently trapped surface, with negative mean normal curvature. 

\section{Application to theorems}

In this section we will abandon the example of quantum energy inequalities. Instead we proceed in a general way, not assuming that the NCC holds. Of course to have exact results we need some form of energy condition and \eqref{eqn:qeiinsp} is a candidate, but perhaps not the only one.

Turning back to the condition for the formation of a focal point \eqref{eq:fp-criteria}, we note that the condition is satisfied if 
	\be
 \label{eqn:suftrapped}
		U_{\mu}\mathrm{H}^{\mu} \big|_{P} < -\frac{1}{n - 2} \inf_{\substack{f\in C^{\infty}_{1,0}[0, \ell]}}J_{\ell}[f] \,,
	\ee
    where
    \be
J_\ell[f]=\int_0^\ell \left((n-2)f'(\lambda)^2-f(\lambda)^2 R_{\mu \nu}U^\mu U^\nu  \right) d\lambda \,.
\ee
This is the general definition of a sufficiently trapped surface, that incorporates the definition of the trapped surface introduced by Penrose, and includes the case of Eq.~\eqref{eqn:trapqei} arising from a quantum energy inequality inspired condition.  

This definition is also sufficient to show that this hypersurface is always behind the event horizon (if such exists), without the need of an energy condition. In particular we have the following proposition
\begin{prop}
\label{prop:surf}
Let $(M,g_{\mu \nu})$ be a strongly asymptotically predictable spacetime. Let $T$ a co-dimension-$2$ spacelike hypersurface with mean normal curvature $H^\mu$ satisfying Eq.~\eqref{eqn:suftrapped}. Then $T \subset B$, where $B$ is the black hole region of spacetime.
\end{prop}
The proof of this proposition is given in Ref.~\cite{Kontou:2023ntd}.

\begin{wrapfigure}{l}{0.4\textwidth}
\includegraphics[width=0.9\linewidth]{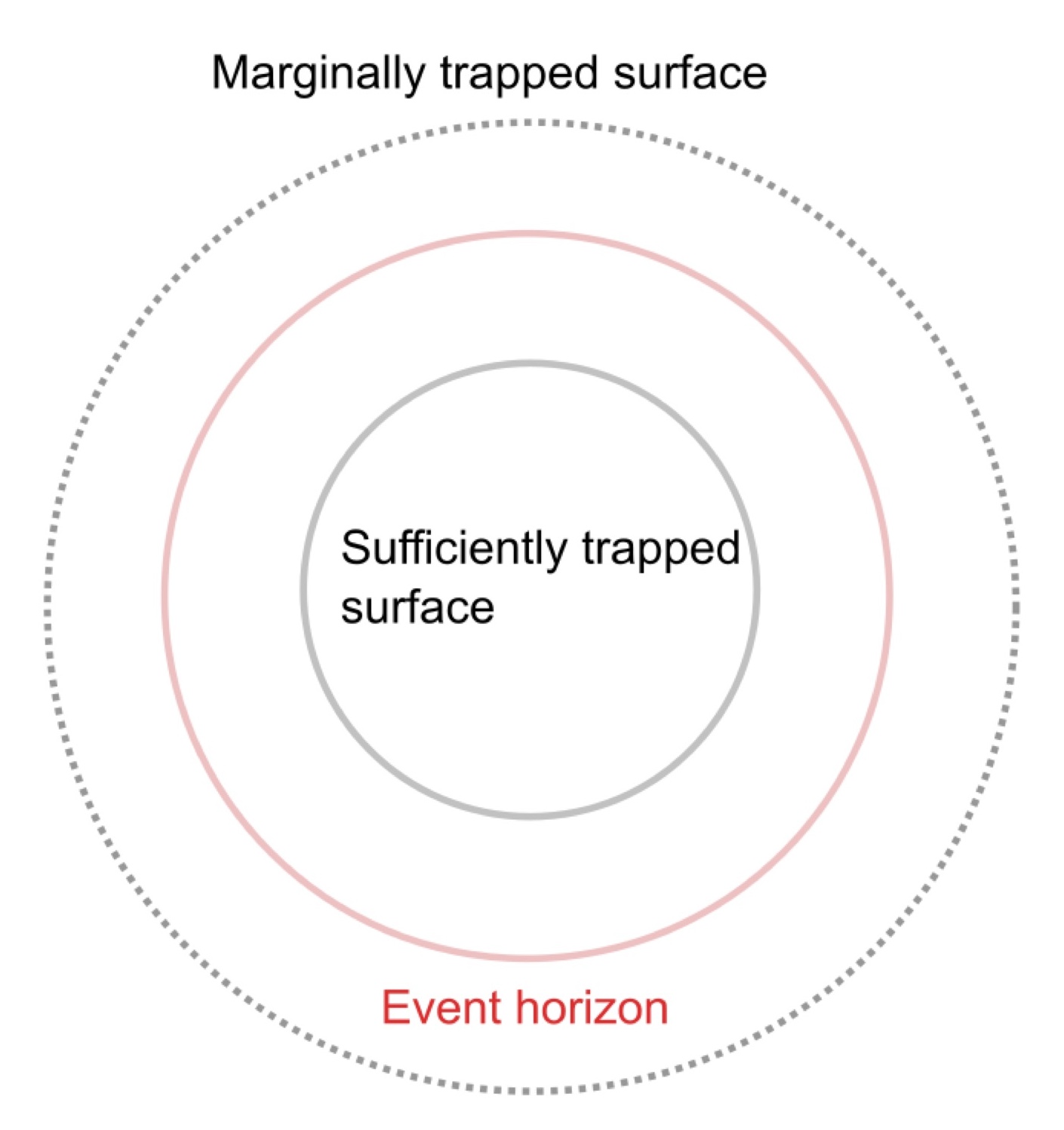} 
\caption{The location of the sufficiently trapped surface in an evaporating black hole spacetime.}
\label{fig:wrapfig}
\end{wrapfigure}

Further, this condition is sufficient to prove null geodesic incompleteness along with the right causality conditions of the spacetime \cite{Fewster:2019bjg}. Indeed, if we consider null geodesics emanating normally from $T$ and condition \eqref{eqn:suftrapped} holding, we have a focusing theorem for the null geodesic congruence. If additionally, our spacetime is globally hyperbolic with non-compact Cauchy hypersurfaces, then it is geodesically incomplete. Thus we can prove a singularity theorem assuming we have sufficiently trapped surface for our energy condition.

Combined with proposition \ref{prop:surf} we have a statement about weak cosmic censorship: The sufficiently trapped surface which leads to null geodesic incompleteness is always behind the event horizon. Thus the singularity is always behind the event horizon if one forms. This applies to evaporating black holes which can have the trapped surface outside of their event horizon but not the sufficiently trapped surface (see Fig.~\ref{fig:wrapfig}). 

For a strongly asymptotically predictable spacetime, we can also reformulate the area theorem using the concept of the sufficiently trapped surface. In particular, if \(H\) is the black hole horizon, \(U^{\mu}\) the tangent field of its null generators $\gamma(\lambda)$, and \(\mathrm{H}^{\mu}\) the mean normal curvature of \(\mathscr{H} = \Sigma \cap H\), where \(\Sigma\) is a Cauchy surface for the globally hyperbolic region $\tilde{V}$, we have
\be
\label{eqn:horizonch}
		\delta_U\mathcal{A}_{\mathscr{H}} = \int_{\mathscr{H}} \mathrm{H}^{\mu}(p)U_{\mu} \ge - \frac{1}{n - 2}\left(\inf_{\substack{f\in C^{\infty}_{1,0}[0, \ell]}}J_{\ell}[f]\right)\cdot\mathcal{A}_{\mathscr{H}}.
\ee
So the area of the horizon is allowed to decrease but no more than what is allowed by the infimum of $J_\ell[f]$, the same quantity that specifies the sufficiently trapped surface.

Finally, the sufficiently trapped surface can be used to prove a version of the Penrose inequality \eqref{eqn:PI} for evaporating black holes. There are two main modifications compared to the original derivation: (i) since the black hole is evaporating there is no end state and instead two initial data sets at affine parameters $0$ and $s$ are compared (ii) the trapped surface is replaced by the sufficiently trapped surface $T$. Then for spherically symmetric initial data\footnote{The initial data is required to satisfy a further condition described in \cite{Hafemann:2025mjf}.}
\be
\label{eqn:PIevap}
m_{\mathrm{ADM}} \geq \sqrt{\frac{A_{\min} (T)}{ 16\pi }} \exp \left(-\frac{1}{4}\int_{0}^s \nu (\lambda)d\lambda \right).
\ee
where $A_{\min} (T)$ is the minimum area required to enclose $T$ and $\nu (\lambda):=\inf_{f \in C_{1, 0}^{\infty} [\lambda, \ell]} J_{\ell} [ f ]$. 

This condition could be used as a test for the existence of an event horizon (as a consequence of the causal structure of spacetime). The concept is as follows: one could examine a black hole model that includes a trapped surface leading to a singularity. Now an observation of violation of the inequality of Eq.~\eqref{eqn:PIevap} with an appropriate quantum energy inequality (or any energy condition), would lead to the conclusion of possible non-existence of an event horizon. This follows as any sufficiently trapped surface with the correct causal structure would be behind the event horizon.

\medskip

In conclusion, the notion of a sufficiently trapped surface provides a robust generalization of the classical trapped surface in regimes where the null convergence condition is violated. The sufficiently trapped surface retains much of the utility of the original concept: it can be employed to establish singularity theorems, derive a modified area theorem, and formulate an appropriate version of the Penrose inequality. Crucially, sufficiently trapped surfaces remain confined within the event horizon, thereby offering significant support for the validity of the weak cosmic censorship conjecture in semiclassical gravity.

\newpage 

\bibliographystyle{plain}
\bibliography{biblio}

\end{document}